\documentclass[
]{ceurart}

\usepackage{listings}
\begin{document}

\copyrightyear{}
\copyrightclause{Copyright for this paper by its authors.
  Use permitted under Creative Commons License Attribution 4.0
  International (CC BY 4.0).}

\conference{}

\title{Supporting the understanding of ontologies for scientific knowledge graphs with the new version of LODE}

\author[1]{Valentina Pasqual}[%
orcid=0000-0001-5931-5187,
email=valentina.pasqual2@unibo.it,
url=https://www.unibo.it/sitoweb/valentina.pasqual2,
]
\cormark[1]

\author[2]{Ahmadreza Nazari}[%
orcid=0009-0000-3884-1575,
email=ahmadreza.nazari@studio.unibo.it
]

\author[1,3]{Silvio Peroni}[%
orcid=0000-0003-0530-4305,
email=silvio.peroni@unibo.it,
url=https://www.unibo.it/sitoweb/silvio.peroni,
]

\address[1]{Digital Humanities Advanced Research Centre (/DH.arc), Department of Classical Philology and Italian Studies, University of Bologna, Bologna, Italy}

\address[2]{Digital Humanities and Digital Knowledge, Department of Classical Philology and Italian Studies, University of Bologna, Bologna, Italy}

\address[3]{Research Centre for Open Scholarly Metadata, Department of Classical Philology and Italian Studies, University of Bologna, Bologna, Italy}

\cortext[1]{Corresponding author.}

\begin{abstract}
    Scientific Knowledge Graphs (SKGs) rely on increasingly complex and heterogeneous semantic models, whose reuse and interoperability require ontologies and other semantic artefacts to be understandable by both machines and humans. However, existing ontology documentation tools provide limited support for modern reuse-oriented modelling practices, where entities are distributed across imported modules and enriched through cross-module annotations. This paper presents the new version of LODE, a re-engineered and extensible framework for generating human-readable documentation of semantic artefacts. LODE preserves the established single-page, W3C Recommendation-style documentation paradigm of its legacy predecessor, and introduces a modular architecture that separates artefact interpretation (Reader), internal representation (Model), and rendering (Viewer). Served as a web service, LODE supports the documentation of OWL ontologies and enables additional documentation features (e.g., entity stand-alone documentation, RDF entity provenance, Markdown rendering). We demonstrate the applicability of LODE through the documentation of the Scientific Knowledge Graph Interoperability Framework ontology (SKG-O), showing how it improves the accessibility and reuse of modular scientific knowledge graph models.
\end{abstract}


\maketitle

\section{Introduction}
\label{sec:introduction}

In 2009, in two seminal articles, David Shotton introduced the concept of \emph{semantic publishing} as ``anything that enhances the meaning of a published journal article, facilitates its automated discovery, enables its linking to semantically related articles, provides access to data within the article in actionable form, or facilitates integration of data between articles''~\cite{shotton_adventures_2009}, which can be reached by ``enriching the article with appropriate metadata that are amenable to automated processing and analysis, allowing enhanced verifiability of published information and providing the capacity for automated discovery and summarization''~\cite{shotton_semantic_2009}. Starting from those bases, the Semantic Web community interested in the scholarly publishing domain has largely investigated and expanded upon such a topic in a plethora of events, often organised in the context of primary research venues such as the European Semantic Web Conference (with the 2011-2014 editions of the SePublica workshop~\cite{garcia_castro_semantic_2014}), the Web Conference (with the 2015-2018 editions of the SAVE-SD workshop~\cite{gonzalez-beltran_semantics_2018}), and the International Semantic Web Conference (with the 2021-2026 editions of the Sci-K workshop~\cite{jacyszyn_scientific_2025}).

Among the various aspects that characterise semantic publishing, ontologies, vocabularies, taxonomies, and other semantic artefacts play a prominent role in implementing a specific data model for the scholarly domain and, thus, in enabling semantic interoperability among systems~\cite{corcho_eosc_2021,nyberg_akerstrom_developing_2024,baumann_landscape_2026}. Indeed, in the past, several ontologies have been proposed, such as the \emph{Semantic Publishing And Referencing (SPAR) Ontologies} (\url{https://www.sparontologies.net})~\cite{rutkowski_spar_2018} and the \emph{Bibliographic Ontology} (BIBO, \url{https://www.dublincore.org/specifications/bibo/}), and such technologies have been adopted by several providers of scientific knowledge graphs (SKGs) such as OpenCitations (\url{https://opencitations.net})~\cite{peroni_opencitations_2020} and Open Research Knowledge Graph (ORKG, \url{https://orkg.org/})~\cite{auer_improving_2020}.

However, to foster broader adoption of such semantic artefacts, an important tool to provide for accompanying an ontology is \emph{human-readable documentation} that enables an easier comprehension of the various types of entities (classes, properties, individuals, etc.) an ontology introduces in a way that is understandable by non-technical experts rather than solely knowledge engineers and semantic web developers. This is even more important when an ontology is defined by gathering together entities from existing models without redefining their original semantics, thus performing a \emph{direct reuse} of ontological terms~\cite{cota_landscape_2020}. However, such a direct reuse can be performed by either \emph{importing} the desired ontologies into a new one (via \texttt{owl:imports} axioms) or including only the selected ontology terms in a new ontology \emph{referencing} the reused ontologies via \texttt{rdfs:isDefinedBy} axioms. While the former case includes the semantics of reused ontologies entirely within the new one, in the latter case the semantics of reused terms are delegated to external ontologies, avoiding the unnecessary inclusion of useless ontological entities. In these cases, the rich visualisation of annotation information for each entity in an appropriate ontology documentation is crucial as it may clarify how the various entities are used in the context of the application.

Several tools, such as LODE~\cite{peroni_live_2012, peroni_tools_2013} and WIDOCO~\cite{garijo_widoco_2017}, have been developed to create ontology documentation, often in HTML, by processing an ontology and its annotations\footnote{Indeed, it is worth mentioning that the first alpha (and internal) version of LODE was created to quickly obtain human-readable documentation for the first set of SPAR Ontologies under development between 2010-2011. Having documentation was crucial at that time to organise several meetings with domain experts, who lacked expertise in ontological development, to discuss various aspects of domain modelling, while masking the obscure technicalities introduced by OWL and the formats used to implement such ontologies. In particular, during these meetings, we often gathered additional information that led to slight changes to parts of the modelling, and having an automated process for creating documentation helped speed up discussions with domain experts. From that point, producing a full standalone software (\url{https://web.archive.org/web/20120822012904/http://sourceforge.net/projects/lode/}), also run as a Web service (\url{https://web.archive.org/web/20120402180140/essepuntato.it/lode}), that may help all ontology engineers to simplify the creation of ontology documentation either for developing or publishing purposes was a natural subsequent step.}. However, their development has been guided by requirements for ontology documentation that have been recently expanded, considering the importance that semantic artefacts have today, such as enabling the possibility of rendering the documentation of single entities within an ontology and the fact that the tool should be agnostic and able to represent all the annotations the entities it includes have independently of the particular annotation vocabulary adopted. Thus, such tools should be extended (or new ones should be developed) to address these needs and be continuously adopted within the scholarly domain -- being them used, for instance, also to produce the documentation of the SKG-IF RDA Recommendation~\cite{mannocci_scientific_2026}.

In this paper, to address the aforementioned needs, we introduce (for the first time) a new version of LODE (v2.0.0)~\cite{pasqual_lode_2026}, now maintained by OpenCitations to ensure its code is continuously developed and sustained, and to guarantee the availability of an online service accessible via a browser and a REST API, which is now included in the OpenCitations infrastructure. This new version includes a full re-engineering of the code, now available in Python, and implements several new features to address all the limitations in the old version of the system. To demonstrate its new capabilities in action, we use the SKG-IF Ontology (SKG-O, \url{https://w3id.org/skg-if/ontology/}) as a case study, showing improvements in its documentation using LODE 2.0, which will be adopted to provide the new documentation of SKG-O in the near future.

The rest of the paper is organised as follows. In Section~\ref{sec:related}, we introduce the most prominent related tools for the production of ontology documentation. In Section~\ref{sec:internal_architecture}, we present the new architectural organisation of LODE 2 and its main components, while in Section~\ref{sec:features} we present all the new features introduced in this version. Finally, in Section~\ref{sec:casestudy}, we show some of the new features of LODE in action during the generation of the documentation for SKG-O, and in Section~\ref{sec:conclusions}, we conclude the paper by sketching out some future work.

\section{Related Work}
\label{sec:related}


SKGs are typically designed and evolved to address specific research objectives and application scenarios. This use-case-driven development results in heterogeneous semantic models, domain-specific vocabularies, and project-dependent modelling decisions, ultimately leading to fragmentation across the SKG ecosystem~\cite{mannocci_scientific_2026}. A recent effort to overcome this fragmentation is the Scientific Knowledge Graph - Interoperability Framework (SKG-IF, \url{https://skg-if.github.io})~\cite{mannocci_scientific_2026}, which provides a shared interoperability layer through common concepts and mappings while preserving the autonomy of individual data models. Its conceptual model is formalised by the SKG-IF Ontology (SKG-O, \url{https://w3id.org/skg-if/ontology/}), thereby enabling interoperability and promoting the long-term sustainability of scientific knowledge infrastructures. Nevertheless, the practical reuse of the framework also relies on making its semantic artefacts understandable to human users.

Among currently maintained ontology documentation approaches, the \emph{Live OWL Documentation Environment} (LODE, \url{https://w3id.org/lode/}) \cite{peroni_live_2012,peroni_tools_2013,peroni_httpsgithubcomessepuntatolode_2020} established the reference model for human-readable ontology documentation and became a \textit{de facto} standard for OWL ontology publication. Introduced in 2012, LODE defined the widely adopted single-page, W3C Recommendation-style representation of ontologies, automatically extracting classes, object and data properties, named individuals, annotation properties, general axioms, and namespace declarations, and presenting them through structured HTML documentation with their textual descriptions and hyperlinks. Built in Java and available on GitHub (\url{https://github.com/essepuntato/lode}), it linearises the ontology to RDF/XML via the OWL API ({\url{https://github.com/owlcs/owlapi})~\cite{palmisano_httpsgithubcomowlcsowlapi_2024} and then applies large XSLT 2.0 stylesheets (\url{https://www.w3.org/TR/xslt20/})~\cite{kay_xsl_2021} to produce HTML documents; optional parameters (e.g., \texttt{owlapi}, \texttt{imported}, \texttt{closure}, \texttt{reasoner}, \texttt{lang}) extended this pipeline. Its adoption by initiatives such as the SPAR Ontologies~\cite{rutkowski_spar_2018}, together with its integration into publication workflows based on content negotiation for dereferenceable ontology URIs, further consolidated the use of automatically generated, single-page HTML documentation as a common practice for Semantic Web publishing. Its monolithic, XSLT-centred design, however, tightly coupled ontology parsing and graphical rendering, making the system hard to extend or modernise after a decade of technological change.

Over the years, LODE has served as the foundation for several ontology documentation tools, either by being integrated as a documentation component, as in Wizard for Documenting Ontologies (WIDOCO, \url{https://github.com/dgarijo/WIDOCO})~\cite{garijo_widoco_2017,garijo_widoco_2024}, or by being re-engineered into new implementations, as in pyLODE (\url{https://github.com/rdflib/pyLODE})~\cite{car_httpsgithubcomrdflibpylode_2026}. pyLODE provides a Python-based reimplementation built on RDFLib (\url{https://github.com/rdflib/rdflib})~\cite{krech_rdflib_2026}, removing the original LODE's Java/XSLT dependency and integrating ontology documentation generation into Python-oriented semantic workflows through a command-line interface, a Python library, a Docker image, and a web service (\url{https://tools.kurrawong.ai/pylode}). However, Abox definition (i.e., named individuals) is not included, and additional functionalities provided by LODE (e.g., imported ontology handling) are not supported.

WIDOCO, by contrast, extends the original LODE pipeline with a broader documentation workflow. It automatically enriches the generated documentation with ontology metadata and publication-oriented features, including interactive visualisation through WebVOWL (\url{https://github.com/visualdataweb/webvowl})~\cite{lohmann_webvowl_2015}, provenance metadata based on PROV-O, licence information, Schema.org annotations, change logs, and content-negotiation support. Additionally, WIDOCO encourages ontology publishers to supply a rich set of metadata to maximise the completeness and quality of the resulting documentation (\url{https://github.com/dgarijo/Widoco/blob/master/doc/metadataGuide/guide.md}) and best practices (\url{https://dgarijo.github.io/Widoco/doc/bestPractices/index-en.html}), and provides rich-text annotations (Markdown rendering facilities). Like pyLODE, however, it produces a single, self-contained HTML documentation page for each ontology. WIDOCO has been used to generate the published documentation of SKG-O \cite{mannocci_scientific_2026}.


\section{Architecture}
\label{sec:arch}

The new version of LODE re-implements existing legacy features (e.g. OWL documentation on single-page and W3C-styled documentation, imported and closure handling, etc.), and extend them in a modernised architecture, replacing the XSLT pipeline with a modular Python-based environment.  While the original LODE was conceived for documenting OWL ontologies, the contemporary Semantic Web landscape increasingly relies on a broader ecosystem of semantic artefacts, including RDF(S) vocabularies, SKOS concept schemes, SHACL shapes, and other knowledge representation resources. Accordingly, the new version of LODE is designed as an extensible documentation framework. OWL documentation represents the first fully supported artefact family, while the architecture is explicitly designed to accommodate additional semantic resource types through future extensions.

\subsection{Internal architecture}
\label{sec:internal_architecture}
The re-engineered version of LODE is designed around a strict separation of concerns, decoupling the system into three layers, as shown in Figure~\ref{fig:arch}. The \emph{Reader} parses the input file that defines the semantic artefact for which the documentation is to be produced, and maps it onto the internal model. The \emph{Model} organises and collects the parsed information into Python classes, providing a uniform and independent layer through which the Reader communicates with the \emph{Viewer}, which uses the populated model to generate the final HTML documentation.


\begin{figure}
  \centering
  \includegraphics[width=\linewidth]{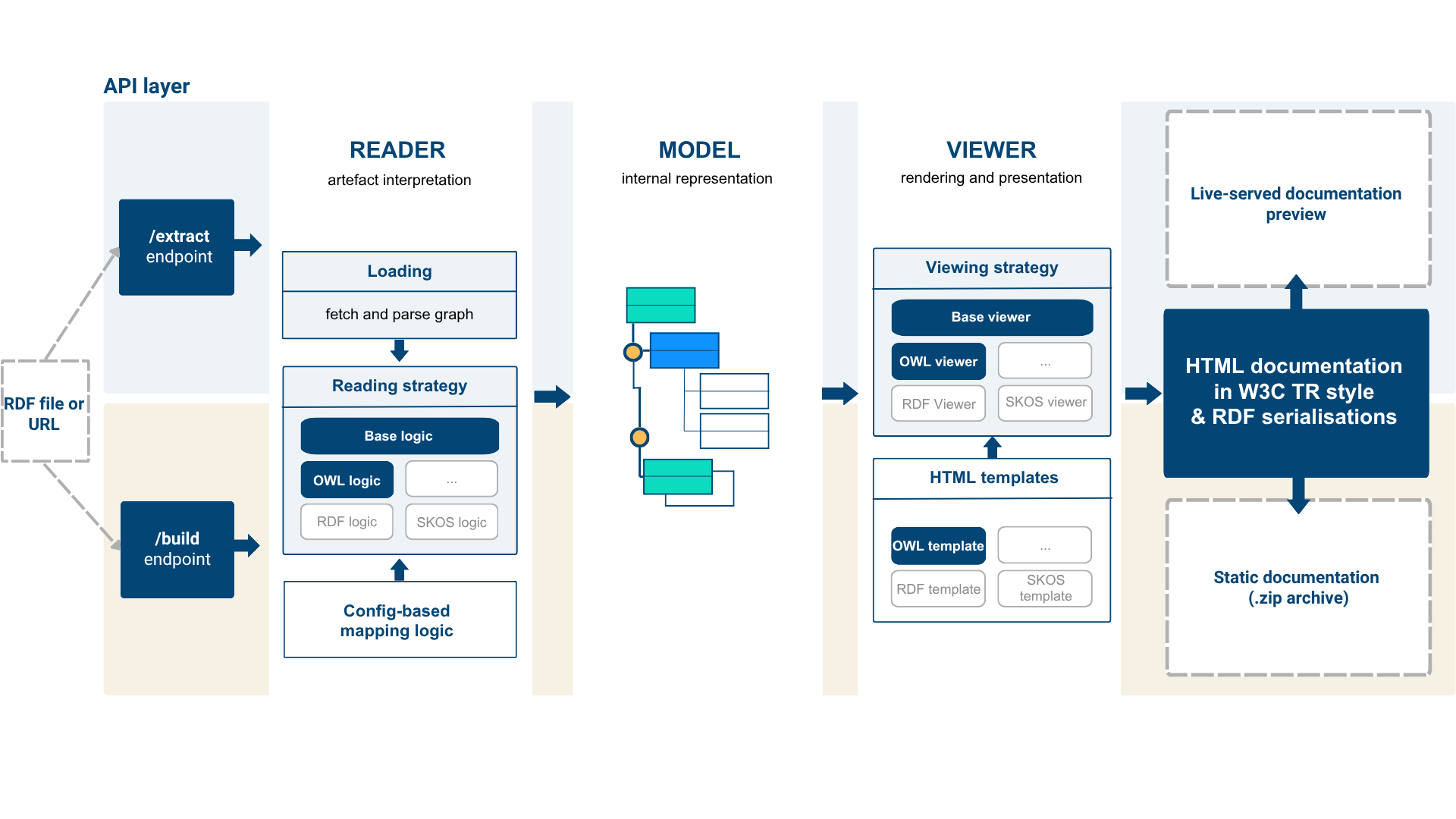}
  \caption{High-level representation of LODE pipeline}
  \label{fig:arch}
\end{figure}

\paragraph{Reader.} The Reader takes as input a semantic artefact (available in RDF) via its URL or file upload (through the \texttt{loader} module) and loads it into an RDFLib graph. A \emph{configuration manager} instantiates the logic layer responsible for interpreting the graph. The framework first provides a base logic that implements the processing of core RDF and RDFS constructs, which are common to semantic artefacts represented as RDF graphs. On top of this layer, family-specific logic modules introduce the additional constructs required by a given semantic artefact model. 

For OWL artefacts, \texttt{OwlConfigManager} selects the OWL configuration and instantiates an \texttt{OwlLogic}, which extends the base logic with OWL-specific processing capabilities. The logic iterates over the input graph and maps triples onto the Model. This mapping is driven by a declarative configuration file that enumerates the vocabulary terms of the family and specifies, for each term, the model class to instantiate and how to populate it. The OWL implementation covers the terms of the OWL namespace (\url{http://www.w3.org/2002/07/owl\#}) following the constructs described in~\cite{world_wide_web_consortium_owl_2012}. 

Declarative mapping is supported by a set of \textit{handler} functions that resolve ambiguous cases from context and perform lightweight, documentation-oriented inferences, such as traversing the class hierarchy to resolve undeclared domains and ranges and assigning default values if nothing is retrieved (e.g. \texttt{owl:Thing}). When a predicate may map to more than one target class, a handler selects the target from context: for instance, \texttt{rdfs:domain} and \texttt{rdfs:range} are dispatched through dedicated handlers that also infer the kind of property involved (object, data or annotation) from how it is used if not explicitly declared in the ontology. Handlers likewise classify blank nodes according to their context, e.g. by classifying restrictions or reified annotations. Additionally, triples not covered by any configuration term are preserved as generic statements to ensure that no assertion is silently dropped (e.g. SKOS annotations). Throughout, the Reader records RDF provenance by locally storing a Model instance containing the set of triples that produced it from the source graph.

\paragraph{Model.} The Model is a unified, extensible object-oriented schema of Python classes, as illustrated in Figure~\ref{fig:model}. Designed as a documentation-oriented representation, it is intended to support faithful documentation and \emph{not} intended as a meta-model of the Semantic Web nor as an alternative to OWL/RDF(S) semantics. 
It is conceived as a shared vocabulary that can represent different types of semantic artefacts through dedicated extensions. The current implementation focuses on representing RDF/RDFS and OWL constructs, corresponding to the base layer (yellow) and the OWL-specific extension layer (green) shown in the UML diagram in Figure~\ref{fig:model}. Additional artefact-specific extensions (e.g. SKOS) are planned to be integrated within the same framework.

The structure shared by all entities (e.g. identifier, deprecation status, labels, notes, definitions, examples and provenance metadata) is represented by the top class \texttt{Resource}. Class-like resources are represented by \texttt{Concept}, which carries the subclass, equivalence and disjointness relations together with the SKOS-style mappings (broad, narrow, related, exact and close match). Properties are represented by \texttt{Property}, holding domain, range and functionality, and specialised into \texttt{Relation} for OWL object properties, \texttt{Attribute} for data properties, and \texttt{Annotation} for annotation properties. Named entities are represented by \texttt{Individual}, while ontology-level metadata (imports, version, prior versions, compatibility) are represented by \texttt{Model}; and literal values with their language and type are represented by \texttt{Literal} and \texttt{Datatype}. Complex OWL~2 class expressions are represented by a hierarchy of restriction classes rooted in \texttt{Restriction}: existential and universal restrictions by \texttt{Quantifier}, cardinality constraints by \texttt{Cardinality}, \texttt{owl:hasValue} restrictions by \texttt{Value}, boolean combinations by \texttt{TruthFunction}, enumerations by \texttt{OneOf} and datatype facets by \texttt{DatatypeRestriction}. SWRL rules are represented by \texttt{Rule}, \texttt{Atom} and \texttt{Variable}. Generic \texttt{Statement}s represent reification and unmapped (possibly reified) statements.

\begin{figure}
  \centering
  \includegraphics[width=\linewidth]{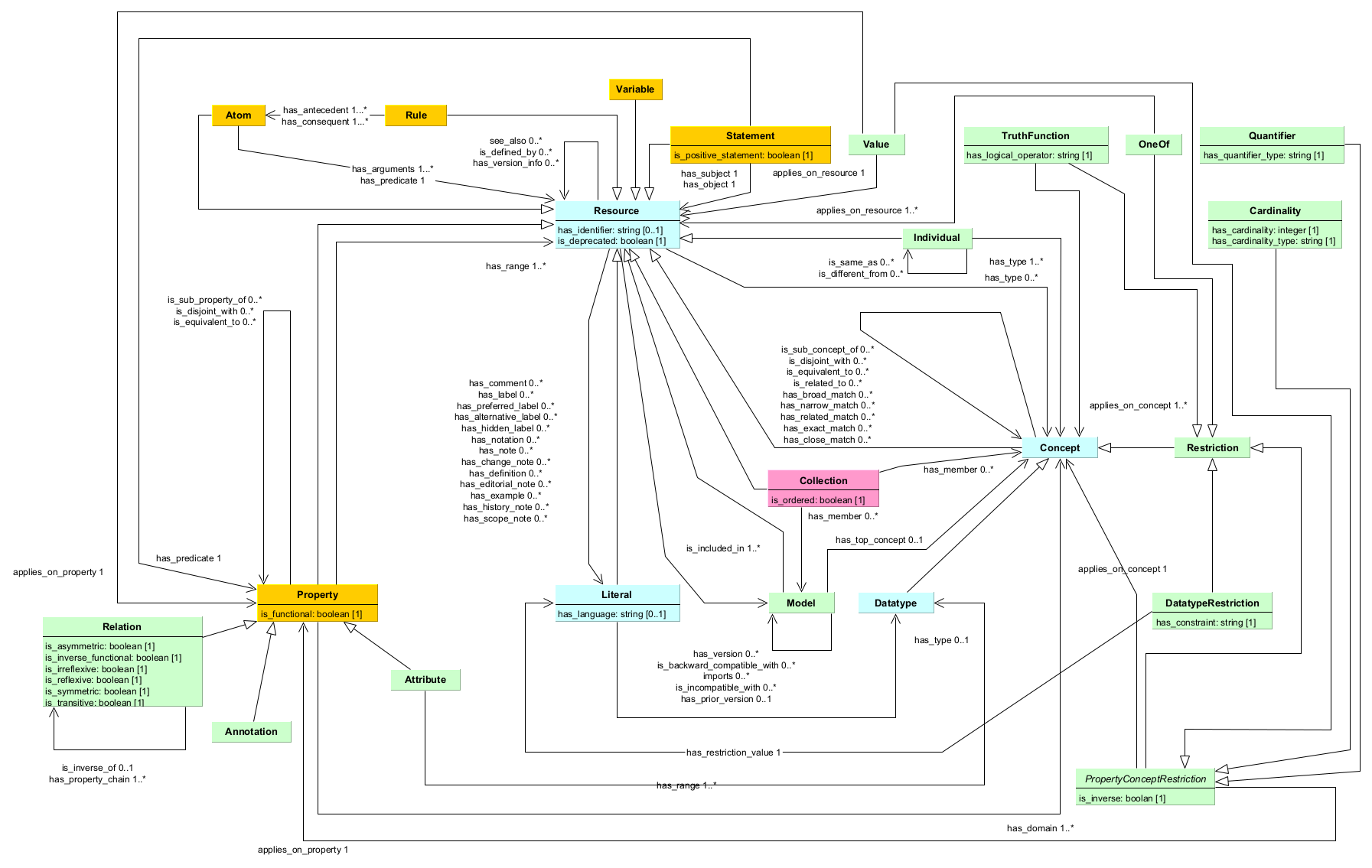}
  \caption{The UML diagram of the Python classes implementing the Model module. In particular, the classes in yellow have been derived from RDF, those in green from RDFS, while the classes in green and pink have been defined from OWL and SKOS, respectively.}
  \label{fig:model}
\end{figure}

\paragraph{Viewer.} The Viewer takes the populated Model produced by the Reader and, accessing it through its defined interface (e.g. OWL), retrieves entities and restructures the information it requires for the sake of presentation. Specifically, it orders entities deterministically so that the same artefact always yields the same output, resolves language tags to select the appropriate label and comment for the target language, and sanitises the Markdown embedded in annotations before rendering it to HTML. The Viewer further implements a client-side rendering pipeline for annotation values containing Markdown. The pipeline detects and transforms Markdown constructs (including clickable URLs, nested lists, and isolated code blocks) into HTML. The generated markup is sanitised using \texttt{marked.js} (\url{https://github.com/markedjs/marked}) \cite{markedjs}
and \texttt{DOMPurify} (\url{https://www.npmjs.com/package/dompurify}) \cite{dompurify},  preventing malformed or potentially unsafe annotations from affecting the documentation structure or client-side execution. In addition, it categorises entities hierarchically, reconstructing the subclass and sub-property trees. The \texttt{OwlViewer} adds the OWL-specific presentation: the entity categories that structure the documentation (concepts, objects, data and annotation properties, and individuals) and their order, reproducing the layout of the legacy LODE. The resulting representation is injected into a Jinja \cite{jinja2} template (\url{https://github.com/pallets/jinja}), styled with Bootstrap 5 \cite{boostrap5} (\url{https://github.com/twbs/bootstrap}) for responsive layout, and uses CSS stylesheets (mimicking the legacy LODE). The Viewer supports two rendering modes: a complete single-page document listing all entities of the artefact (as done in the legacy LODE), and a per-resource view that documents an individual entity on a dedicated page.

\subsection{Deployment, maintenance and hosting}
\label{sec:deploy}
LODE~2 is available at \url{lode.opencitations.net} (and accessible at \url{https://w3id.org/lode} via redirection). The service is exposed as a FastAPI\footnote{\url{https://fastapi.tiangolo.com/}} service through two endpoints:

\begin{itemize}
    \item \texttt{/extract} renders a single semantic artefact on demand. It accepts the semantic artefact (via \texttt{url} or file upload), a \texttt{read\_as} semantic artefact type (e.g. \texttt{owl} for an OWL ontology), an optional \texttt{resource} for single-entity mode, a \texttt{lang} filter, and the \texttt{imported}/\texttt{closure} switches inherited from legacy LODE. Through HTTP content negotiation, the same request returns either the rendered HTML documentation or a machine-readable serialisation (Turtle, RDF/XML, JSON-LD and N-triples), for the whole ontology (including the full set of ontology axioms) or for the selected resource (sub-graph containing all axioms explicitly asserted about that entity).
    \item \texttt{/build} produces a self-contained static documentation site, packaged as a \texttt{.zip} archive, for offline browsing or self-hosting. For automated and reproducible workflows (e.g. CI/CD publishing to GitHub Pages), the same static site is generated by the \texttt{lode build} command-line tool.
\end{itemize}

Critically for a scholarly-infrastructure tool, long-term maintenance and hosting are undertaken by OpenCitations -- one of the infrastructures implementing SKG-IF. Persistent \texttt{w3id.org} URIs with content negotiation are being provisioned, so that ontology URIs can dereference directly to their LODE~2 documentation.

\section{New Features}
\label{sec:features}

The new version of LODE preserves the original single-page, W3C Recommendation-style documentation paradigm, including support for imported ontologies and their closure, while extending its capabilities to support modern semantic artefacts (e.g. including user-friendly navigation for large imported ontologies and distributed annotations handling, as shown in Section \ref{sec:casestudy}). The main new features introduced by LODE can be summarised as follows.

\paragraph{Improved documentation generation and completeness.} Mechanisms that enhance the quality and coverage of the generated documentation:

\begin{itemize}
    \item \textit{Documentation-oriented inference and default values.}
    A lightweight documentation-oriented inference mechanism infers missing information and applies default values to improve documentation completeness. Missing property types (object, datatype, or annotation properties) are inferred from property hierarchy relations, inverse property axioms, and RDF usage patterns (i.e. whether a property connects resources or literals). When domain or range axioms are not explicitly declared, the system recovers inherited constraints by traversing the property hierarchy and applies OWL-compatible defaults when no information is available (i.e. \texttt{owl:Thing} for object property domain and range, and \texttt{rdfs:Literal} for data property range). If no further declaration or usage pattern is available, properties are conservatively classified as annotation properties.

    \item \textit{Safe rich-text annotations rendering.} Markdown content is consistently rendered in all textual literals, independently of the annotation property through which they are provided. LODE adopts an annotation-agnostic approach to preserve documentation-oriented formatting, including headings, inline and block code, bullet lists, hyperlinks, and other standard Markdown constructs.
\end{itemize}

\paragraph{Enhanced entity-level documentation.} Mechanisms that enrich the documentation provided for the documentation of individual entities:

\begin{itemize}
    \item \textit{Dedicated Statements block.}
    Building upon the entity cards introduced in its legacy version, which organises the documentation of each entity around its extracted label (displayed as the card title), IRI, and OWL axioms, each card is extended by a dedicated \emph{Statements} section. This section exposes RDF(S)/OWL assertions that cannot be represented through predefined documentation templates. This prevents information loss by preserving entity annotations expressed through arbitrary properties, including those not explicitly supported by the platform's rendering rules, keeping them graphically separated from formal OWL axioms.

    \item \textit{Per-entity RDF provenance.}
    Each entity card includes a collapsible \emph{RDF Provenance} panel exposing the RDF description underlying the documented resource. By providing direct access to the associated triples in multiple serialisations (Turtle, RDF/XML, JSON-LD, and N-Triples), it enables source-level verification at entity granularity, allowing users to validate individual documentation entries without navigating the entire documentation.

    \item \textit{Stand-alone entity documentation page.}
    Each explicitly defined entity in the ontology is associated with a dedicated link next to its IRI that links to a stand-alone documentation page for that entity. The page supports the same content-negotiation mechanisms as the full documentation and exposes the RDF sub-graph centred on the selected entity, i.e. the same description provided in the \emph{RDF Provenance} panel. This enables dereferenceable, entity-level documentation, where the documented scope is the entity itself rather than the complete ontology.
\end{itemize}

\paragraph{Enhanced navigation and entity discovery.} Mechanisms supporting the exploration of large documentation outputs:

\begin{itemize}
    \item \textit{Structural navigation aid.} A persistent, collapsible sidebar reproduces the complete structure of the ontology table of contents, exposing all documented sections' respective entities through a listed view to aid browsing and discovery across large documentation outputs.
        
    \item \textit{Client-side search.} An embedded, persistent search bar enables direct retrieval of documented entities by matching entity labels, displaying the number of occurrences and dynamically synchronising with the sidebar navigation. A graphical cue highlights matching entities within the documentation body, while non-matching entries in the sidebar are visually de-emphasised.
\end{itemize}

\paragraph{Flexible deployment and accessibility.} Mechanisms supporting reproducible generation and distribution of documentation:

\begin{itemize}
    \item \textit{Local artefact processing.}
    LODE supports the direct upload and processing of local ontology files, extending the previous URL-based workflow of legacy LODE. 

    \item \textit{Static documentation generation.}
    In addition to the live documentation endpoint inherited from its legacy version (\texttt{/extract}), LODE provides dedicated build functionality to generate self-contained static documentation sites. The generated package includes both the complete ontology documentation page and dedicated pages for each explicitly defined entity, preserving the same entity-level views available in the live version. Static sites can be produced through the web service API or the command-line interface, enabling offline browsing, self-hosting, and automated publication workflows (see Section \ref{sec:deploy}).    
\end{itemize}

\section{Case Study: Documenting the SKG-IF Ontology}
\label{sec:casestudy}

As claimed in Section \ref{sec:related}, SKG-O documentation is currently generated using WIDOCO~\cite{garijo_widoco_2017}. We compare the documentation generated for the same ontology using WIDOCO and the new version of LODE, highlighting differences in annotation coverage, rendering strategies, and newly introduced documentation capabilities over the chosen case study.

Figure \ref{fig:comparison-model} compares the current SKG-O documentation generated with WIDOCO (left) with the documentation produced by LODE (right). The WIDOCO rendering reflects the subset of ontology annotations selected through its wizard-driven workflow, as described in Section~\ref{sec:related}. Conversely, LODE adopts an annotation-agnostic rendering approach, exposing the complete set of available ontology-level metadata (e.g. dates, preferred namespace prefixes, and namespace URIs in Figure~\ref{fig:comparison-model}) without requiring these elements to be mapped to predefined documentation templates. 

\begin{figure}
  \centering
   \includegraphics[width=\linewidth]{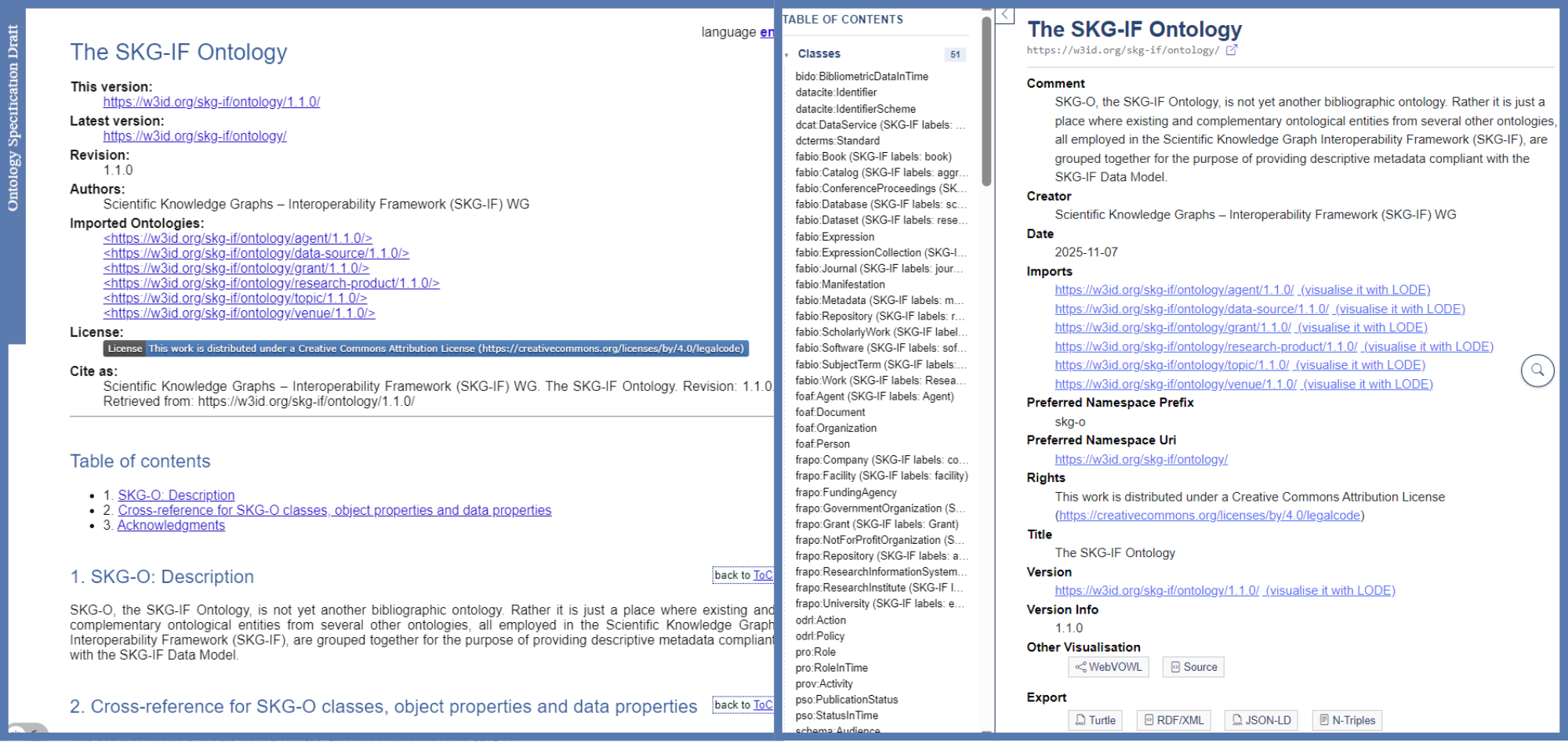}
  \caption{SKG-O ontology documentation generated by WIDOCO (left) and by LODE (right)}
  \label{fig:comparison-model}
\end{figure}

SKG-O adopts a modular architecture in which its six core entities are organised into dedicated ontology modules formally imported by the core ontology, namely agent (\url{https://w3id.org/skg-if/ontology/agent/}), data source (\url{https://w3id.org/skg-if/ontology/data-source/}), grant (\url{https://w3id.org/skg-if/ontology/grant/}), research product (\url{https://w3id.org/skg-if/ontology/research-product/}), topic (\url{https://w3id.org/skg-if/ontology/topic/}), and venue (\url{https://w3id.org/skg-if/ontology/venue/})~\cite{mannocci_scientific_2026}. Each module is documented on a separate page, and users navigate the ontology network via the module URIs provided among the ontology annotations ("Imported Ontologies" in the left panel of Figure \ref{fig:comparison-model}).

LODE supports both distributed and aggregated renderings of documentation. Similarly to WIDOCO, imported ontologies can be documented independently (as in \url{https://lode.opencitations.net/extract?url=https://w3id.org/skg-if/ontology/&read_as=owl}). Alternatively, the \texttt{imported} parameter resolves the direct imports and generates a single documentation page containing approxiately 180 entities, both locally defined and imported (as in \url{https://lode.opencitations.net/extract?url=https://w3id.org/skg-if/ontology/&read_as=owl&imported=true}). Given such an extended documentation, simplified access to documented entities is provided through the sidebar and, additionally, the search bar shown in Figure \ref{fig:comparison-model}. \\


For what concerns entity documentation, we compare the class \textit{fabio:Catalog (SKG-IF labels: aggregator)} as displayed by WIDOCO (Figure \ref{fig:widoco-catalog-entity}) and LODE (Figure \ref{fig:lode-catalog-entity}). 

\begin{figure}
  \centering
  \includegraphics[width=\linewidth]{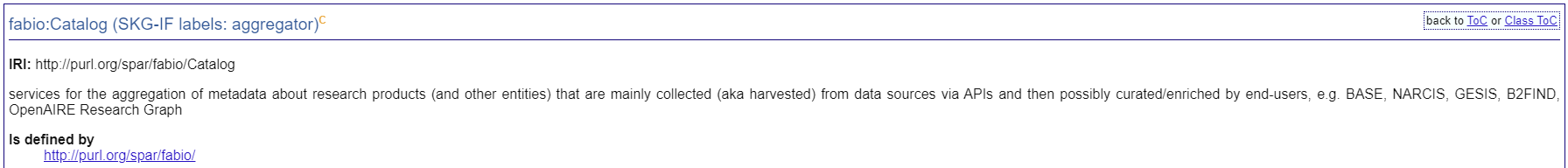}
  \caption{fabio:Catalog (SKG-IF labels: aggregator) class as displayed in SKG-O documentation generated by WIDOCO.}
  \label{fig:widoco-catalog-entity}
\end{figure}

\begin{figure}
  \centering
   \includegraphics[width=\linewidth]{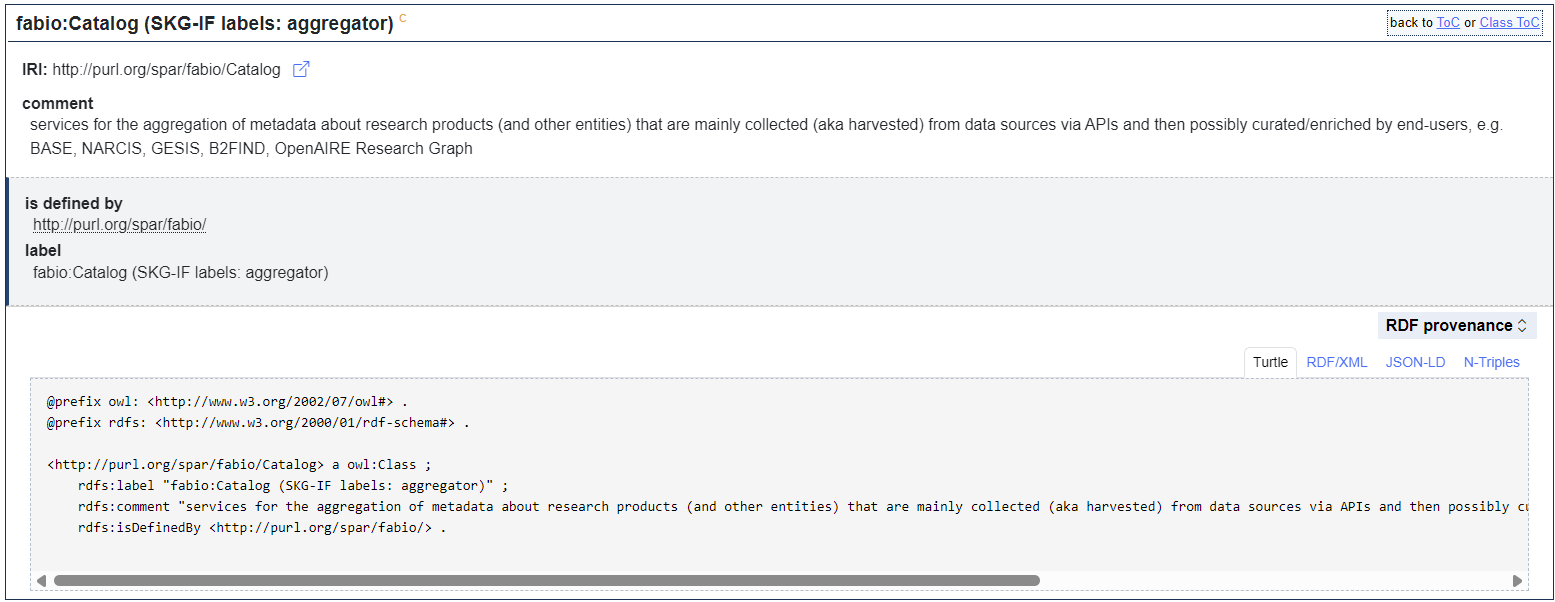}
  \caption{fabio:Catalog (SKG-IF labels: aggregator) class as displayed in SKG-O documentation generated by LODE.}
  \label{fig:lode-catalog-entity}
\end{figure}

In WIDOCO, the class documentation is displayed with the corresponding label as its title (retrieved by \texttt{rdfs:label}), its IRI, its defining ontology (retrieved by \texttt{rdfs:isDefinedBy}) and its comment (retrieved by \texttt{rdfs:comment}). LODE's generated documentation preserves the same information while enriching each entity description with additional documentation-oriented features. These include explicit label metadata (derived from \texttt{rdfs:label} to support multiple labels and multilingual descriptions), an \emph{RDF Provenance} panel exposing the RDF statements used to generate the documentation, with direct traceability to the RDF source, and a direct link (external-link icon next to the entity IRI) to the stand-alone documentation page of the entity. 

As shown in Figure~\ref{fig:lode-catalog-entity-standalone}, the stand-alone documentation page of the entity (\url{https://lode.opencitations.net/extract?url=https://w3id.org/skg-if/ontology/&read_as=owl&imported=true&resource=http://purl.org/spar/fabio/Catalog}) is documented with the complete set of information previously described, while the page additionally provides contextual metadata about the origin of the documented artefact, including the source ontology (\textit{as defined in the SKG-IF ontology}) and its associated metadata, accessible through a dedicated accordion section. Moreover, the documented entity can be retrieved in four RDF serialisations (Turtle, RDF/XML, JSON-LD, and N-Triples) through the dedicated export functionality (\textit{Export entity}).\\

\begin{figure}
  \centering
  \includegraphics[width=\linewidth]{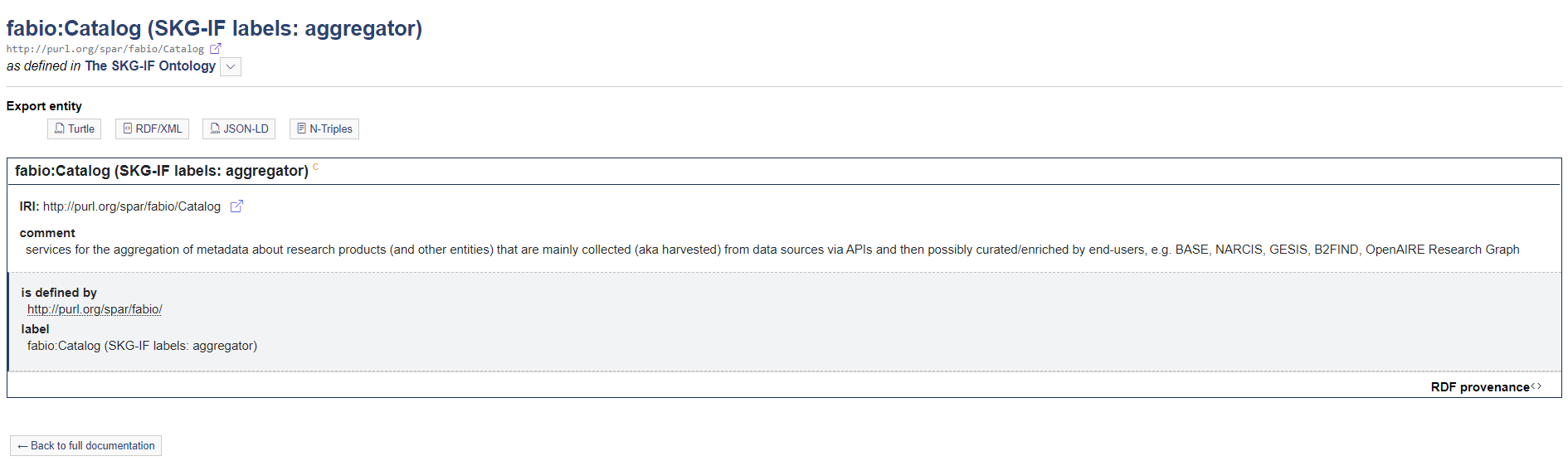}
  \caption{fabio:Catalog (SKG-IF labels: aggregator) class as displayed in SKG-O entity stand-alone documentation generated by LODE.}
  \label{fig:lode-catalog-entity-standalone}
\end{figure}

Additionally, most SKG-O terms are characterised by pervasive reuse: entities are 
frequently borrowed from and re-defined across established vocabularies~ \cite{mannocci_scientific_2026} (e.g. several SPAR ontologies~\cite{rutkowski_spar_2018}). As a consequence, the same entity may be referenced by multiple SKG-O modules, with each module contributing additional annotation assertions that progressively enrich its description. For example, the object property \texttt{foaf:homepage} is reused across different modules (e.g. agent and grant), resulting in a distributed set of annotations associated with the same RDF resource. 


Both WIDOCO and LODE support the inclusion of imported ontologies during documentation generation (through WIDOCO's \texttt{-displayDirectImportsOnly} execution option and LODE's \texttt{imported} parameter). Figures \ref{fig:widoco-homepage-entity} and \ref{fig:lode-homepage-entity} illustrate the documentation of \texttt{foaf:homepage} as generated by WIDOCO and LODE with imported ontologies enabled. 

\begin{figure}
  \centering
  \includegraphics[width=\linewidth]{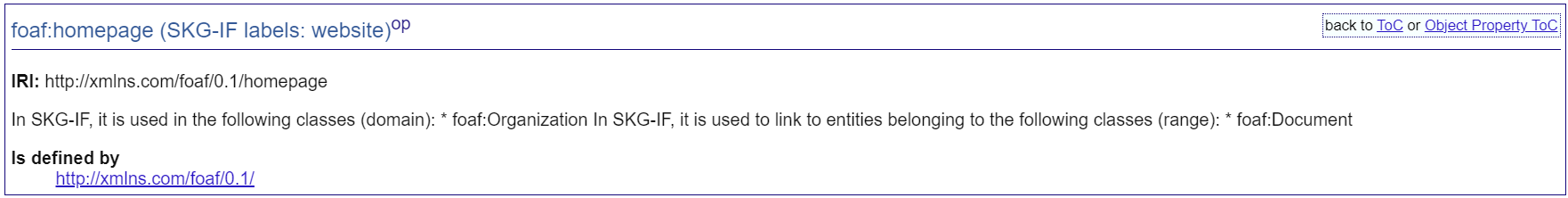}
  \caption{\texttt{foaf:homepage} (SKG-IF label: website) object property as displayed in the SKG-O documentation generated by WIDOCO with imported ontologies enabled.}
  \label{fig:widoco-homepage-entity}
\end{figure}

\begin{figure}
  \centering
   \includegraphics[width=\linewidth]{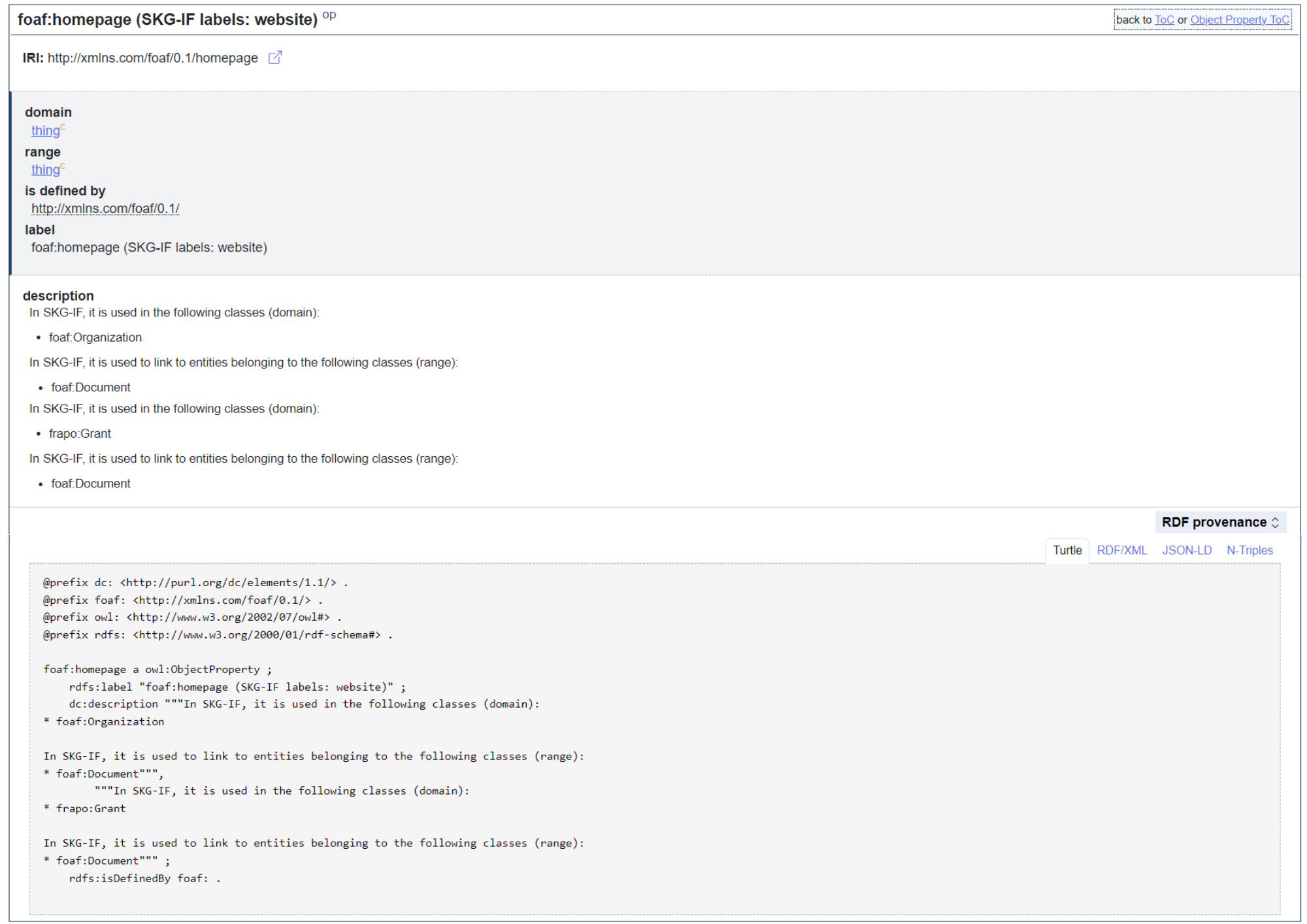}
    \caption{\texttt{foaf:homepage} (SKG-IF label: website) object property as displayed in the SKG-O documentation generated by LODE with imported ontologies enabled.}
  \label{fig:lode-homepage-entity}
\end{figure}

WIDOCO-generated documentation displays the object property with its title (retrieved from \texttt{rdfs:label}), its IRI, its defining ontology (retrieved through \texttt{rdfs:isDefinedBy}), and a textual description. In the latter case, the description is retrieved from \texttt{dc:description}, one of the annotation properties explicitly supported by WIDOCO's documentation mapping. The value contains Markdown syntax, which is displayed verbatim as plain text in the generated documentation. Although \texttt{foaf:homepage} is annotated with different descriptions across multiple SKG-O modules, the generated documentation presents a single description for the imported property.

LODE preserves the information displayed by WIDOCO and enriches the entity documentation with the additional documentation-oriented features, as already described while illustrating the case of the \texttt{fabio:Catalog} class shown in Figure \ref{fig:lode-catalog-entity}.

Additionally, the displayed descriptions include annotations from different ontology modules, which are aggregated into a single entity view through cross-module annotation resolution. LODE does not rely on a predefined annotation vocabulary for textual descriptions; therefore, properties such as \texttt{dc:description}, when not explicitly mapped as documentation fields, are preserved and displayed as general statements (modelled as \texttt{Statement} in LODE's internal model) associated with the entity. This annotation-agnostic approach allows the documentation to expose textual information regardless of the vocabulary used to provide it, supporting future vocabulary changes or newly introduced annotation predicates. Moreover, annotation values are rendered as Markdown whenever applicable, including descriptive literals, preserving the intended formatting of textual documentation. Finally, the entity card is enriched with domain and range information. When explicit axioms are not available, LODE applies documentation-oriented defaults and displays \texttt{owl:Thing} as the inferred domain and range. To preserve transparency between documentation-oriented inferences and axioms explicitly asserted in the ontology, the \emph{RDF Provenance} panel only reports source-level RDF statements; consequently, inferred default values (such as the \texttt{owl:Thing} domain and range shown in this case) are not included among the provenance triples.

\section{Conclusions}
\label{sec:conclusions}
This work presented the new version of LODE, a re-engineered and extensible framework for generating human-readable documentation of semantic artefacts. LODE reimplements the functionalities provided by its predecessor, including single-page W3C Recommendation-style OWL documentation, imported ontologies, and transitive-closure handling. Additionally, it replaces the original XSLT pipeline with a modular Python-based architecture built upon a strict separation of concerns among its three main components (Reader, Model, and Viewer). Currently, OWL is the first fully supported semantic artefact family; however, the decoupled architecture is designed to accommodate additional semantic resource types, such as RDF(S) vocabularies, SKOS concept schemes, SHACL shapes, and other knowledge representation resources, through dedicated extensions.

The current implementation provides full support for OWL artefacts and introduces several improvements aimed at addressing the documentation requirements of modern semantic models. In particular, LODE extends the original documentation workflow by improving (i) documentation generation and completeness (in particular, by providing documentation-oriented inferences and default values and rendering rich-text annotations), (ii) entity-level documentation and traceability (in particular, by adding annotation-agnostic statements, per-entity RDF provenance, and entity stand-alone documentation), (iii) navigation and discovery of documented entities (in particular, by providing a persistent, collapsible sidebar for ontology structural navigation aid and a client-side search bar), and (iv) the generation and deployment of documentation artefacts (in particular, by providing local artefact upload and static documentation generation through the \texttt{/build} endpoint).

The applicability of these features has been demonstrated by comparing the current SKG-O documentation generated with WIDOCO and the documentation produced by the new version of LODE for the same ontology. The comparison highlights differences in annotation coverage, rendering strategies, and newly introduced documentation capabilities. It demonstrates that LODE provides richer documentation outputs by aggregating entity descriptions distributed across imported ontology modules, preserving heterogeneous annotations independently of the adopted vocabulary, rendering rich-text documentation consistently, and exposing the RDF provenance of each documented entity.   

Currently, this new version of LODE is being used to develop the new documentation pages for all the SPAR Ontologies, largely extending their annotations to provide even more self-contained documentation. Similarly, we plan to use LODE to provide the new documentation for SKG-O, thereby making the SKG Data Model clearer for users. 

Additional future work will focus on extending the range of supported semantic artefact families while reusing the existing modular architecture. The first planned extensions concern the introduction of dedicated modules for RDF resources without ontology-level semantics and for SKOS concept schemes, followed by support for SHACL shapes. Moreover, future developments will investigate interactions among different semantic artefact modules, such as the combined documentation of OWL ontologies and SHACL shapes. Finally, we will also work on enabling LODE to use an OWL reasoner to complement its current lightweight, documentation-oriented inference mechanisms.

\paragraph{Acknowledgments.} This work has been funded by the European Union’s Horizon Europe framework programme under Grant Agreement No 101187940 (LUMEN).

\paragraph{Declaration of Use of Generative AI.} No generative AI systems were used to generate or analyse data, or draft the scientific content of this manuscript. AI-assisted tools were used only for spell-checking and minor grammar suggestions on the English text. Generative AI–based coding assistants were used during the development of LODE to support code generation and, especially, review; all AI-suggested code was reviewed, tested, and validated by the authors.

\newpage

\bibliography{references}

\end{document}